\documentclass[aps,prb,floatfix,twocolumn,showpacs,preprintnumbers,amsmath,amssymb,superscriptaddress]{revtex4}

\usepackage{graphicx}  
\usepackage{dcolumn}   
\usepackage{bm}        
\usepackage{stmaryrd}  
\usepackage{multirow}  
\usepackage{color}     
\usepackage{amsmath,amssymb}

\newcommand{\der}[3]{\frac{{\rm d}^{#1} #2}{{\rm d} #3^{#1}}}     
\newcommand{\av}[1]{ {\langle #1 \rangle} }                       

\def   \bhs      {{\hat{\bm s}}}                         
\def   \sx       {{s_{x}}}
\def   \sy       {{s_{y}}}
\def   \sz       {{s_{z}}}
\def   \SL       {{\hat{\bm S}_{\rm L}}}                 
\def   \SR       {{\hat{\bm S}_{\rm R}}}                 
\def   \Ms       {M_{\rm s}}                             
\def   \TOR      {\bm\Gamma}                             
\def   \tor      {\bm\tau}                               

\def   \ts       {t_{\rm s}}                             
\def   \ema      {{\overline{s_{z}}}}                    

\def   \Heff     {{\bm H}_{\rm eff}}                     
\def   \hext     {H_{\rm ext}}                           
\def   \hani     {H_{\rm ani}}                           
\def   \Hdemag   {{\bm H}_{\rm dem}}                     
\def   \DEMAG    {\bar{\bm N}}                           

\def   \ex       {{\hat{\bm e}_{x}}}                     
\def   \ez       {{\hat{\bm e}_{z}}}                     
\def   \ep       {{\hat{\bm e}_{\phi}}}                  
\def   \et       {{\hat{\bm e}_{\theta}}}                

\def   \gyro     {\gamma_{\rm g}}                        

\def   \nm       {{\rm nm}}                              
\def   \cm       {{\rm cm}}                              
\def   \Amp      {{\rm A}}                               

\def   \FL       {{\rm F_L}}
\def   \NL       {{\rm N_L}}
\def   \FC       {{\rm F_C}}
\def   \NR       {{\rm N_R}}
\def   \FR       {{\rm F_R}}

\def  \aL        {a_{\rm L}}
\def  \aR        {a_{\rm R}}
\def  \bL        {b_{\rm L}}
\def  \bR        {b_{\rm R}}
\def  \vp        {v_{\phi}}
\def  \vt        {v_{\theta}}
\def  \torip    {\bm\tau_{\parallel}}
\def  \torop    {\bm\tau_{\perp}}
\def  \torp      {\tau_{\phi}}
\def  \tort      {\tau_{\theta}}
\def  \tortip   {\tau^{\parallel}_{\theta}}
\def  \tortop   {\tau^{\perp}_{\theta}}
\def  \torpip   {\tau^{\parallel}_{\phi}}
\def  \torpop   {\tau^{\perp}_{\phi}}
\def  \Hdx      {H_{{\rm d} x}}
\def  \Hdy      {H_{{\rm d} y}}
\def  \Hdz      {H_{{\rm d} z}}

\begin{document}

\preprint{APS/123-QED}

\title{Current-induced dynamics in non-collinear dual spin-valves}

\author{P.~Bal\'a\v{z}}
\email{balaz@amu.edu.pl}
\affiliation{Department of Physics, Adam Mickiewicz University,
             Umultowska 85, 61-614~Pozna\'n, Poland}
\author{M.~Gmitra}
\affiliation{Institute of Phys.,
             P. J. \v{S}af\'arik University, Park Angelinum 9,
             040 01 Ko\v{s}ice, Slovak Republic}
\author{J.~Barna\'s}
\affiliation{Department of Physics, Adam Mickiewicz University,
             Umultowska 85, 61-614~Pozna\'n, Poland}
\affiliation{Institute of Molecular Physics, Polish Academy of Sciences
              Smoluchowskiego 17, 60-179 Pozna\'n, Poland}

\date{\today}


\begin{abstract}
  Spin-transfer torque and current induced spin dynamics in
  spin-valve nanopillars with the free magnetic layer located between two magnetic films of
  fixed magnetic moments
  is considered theoretically.
  The spin-transfer torque in the limit of diffusive spin transport
  is calculated as a function of  magnetic configuration.
It is shown that non-collinear magnetic configuration of the
outermost magnetic layers has a strong influence on the spin
torque and spin dynamics of the central free layer.
  Employing macrospin simulations we make some predictions on the free layer
  spin dynamics in spin valves composed of various magnetic layers.
  We also present a formula for critical current in non-collinear magnetic configurations,
  which shows that the magnitude of critical current
  can be several times smaller than that in typical single spin valves.
\end{abstract}


\maketitle


\section{Introduction}
\label{Sec:Introduction}

Magnetic multilayers are ones of the most relevant elements in the
development of cutting-edge technologies. If current flows along
the axis of a metallic hybrid nanopillar structure, a spin
accumulation builds up in the vicinity of the
normal-metal/ferromagnet interface. Moreover, the current produces
spin-transfer torque (STT) which acts on magnetic moments of the
ferromagnetic layers \cite{Slonczewski1996:JMMM,Berger1996:PRB}.
As a consequence, magnetization of a particular layer may be
driven to an oscillation mode \cite{Kiselev2003:Nature}, or can be
switched between possible stable states \cite{Katine1999:PRL}.
However, current-induced control of magnetic moments requires a
rather high current density. Therefore, reduction of the critical
current density remains the most challenging requirement from the
point of view of possible applications. Since the critical
currents are related to STT, which in the diffusive transport
regime is proportional to spin accumulation at the
normal-metal/ferromagnet interface, one may alternatively rise a
question of possible ways to enhance the spin accumulation.

One of the possibilities to decrease the critical current density
in metallic structures  is the geometry proposed by
L.~Berger~\cite{Berger2003:JAP}, in which the free magnetic layer
is located between two pinned magnetic layers of opposite
magnetizations. Indeed, in such a {\em dual spin valve} (DSV)
geometry both interfaces  of the free layer can generate STT, and
this may decrease the critical current several
times~\cite{Berger2003:JAP}.

To examine the influence of an additional magnetic layer on the
free layer's spin dynamics one needs to find first the STT acting
on the central layer for arbitrary direction of its magnetization
vector. To do this, we employ the spin-dependent diffusive
transport approach~\cite{Barnas2005:PRB}, based on Valet-Fert
description~\cite{Valet1993:PRB}. In this paper we present a
comprehensive survey of STT-induced effects in DSVs for generally
non-collinear magnetic configuration of the outermost magnetic
films. We also study asymmetric exchange-biased DSV, in which
magnetic moment of one of the outer magnetic layers is fixed  to
an antiferromagnetic layer due to exchange anisotropy. We show
that a non-standard (wavy-like) angular dependence of STT --
originally predicted for asymmetric spin valves
\cite{Gmitra2006:PRL,Barnas2005:PRB} -- can also occur in DSV
geometry. Such a non-standard angular dependence of the torque is
of some importance from the application point of view, as it
allows to induce steady-state precessional modes without external
magnetic field \cite{Gmitra2006:PRL,Boulle2007:NP}. Furthermore,
we examine current-induced spin dynamics within the macrospin
model, and derive a formula for critical currents which
destabilize trivial fixed points of the central spin's dynamics in
non-collinear configurations of  the outermost magnetizations.

In section \ref{Sec:Model} we describe the model and introduce
basic formula for  STT and spin dynamics in DSVs. Then, in section
 \ref{Sec:AntisymGeom} we  study dynamics in a symmetric
DSV geometry, where we compare STT and spin dynamics in DSVs and
single spin valves (SSVs). Finally, in section \ref{Sec:EBDSV} we
analyze an exchange-biased DSV structure with noncollinear
magnetic moments of the outermost magnetic layers. Finally, we
conclude in section \ref{Sec:Conclusions}.


\section{Model}
\label{Sec:Model}

\begin{figure}[!h]
  \centering
  \includegraphics[width=0.75\columnwidth]{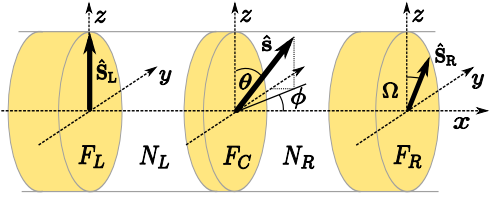}
  \caption{Schematic of a dual spin valve with fixed magnetic moments of
           the outer magnetic
           layers, $\FL$ and $\FR$, and free central magnetic
           layer, $\FC$,
           separated by normal-metal layers $\NL$ and $\NR$. Here, $\SL$, $\SR$, and
           $\bhs$ are unit vectors along the spin moments of the $\FL$, $\FR$, and  $\FC$
           layers, respectively.}
  \label{Fig:DSV_scheme}
\end{figure}

We consider a multilayer nanopillar structure, $\FL / \NL / \FC /
\NR / \FR$, consisting of three ferromagnetic (F) layers separated
by normal-metal (N) layers; see Fig.~\ref{Fig:DSV_scheme}. Spin
moment of the central layer, $\FC$, is free to rotate, while the
right $\FR$ and left $\FL$ ferromagnetic layers are much thicker
and their net spin moments are assumed to be fixed for current
densities of interest. Fixing of these moments can be achieved
either by sufficiently strong coercieve fields, or by exchange
anisotropy at interfaces with antiferromagnetic materials.

In the Landau-Lifshitz-Gilbert phenomenological description, the
dynamics of a unit vector along the net spin moment $\bhs$ of the
central (free) magnetic layer is described by the equation
\begin{equation}
\label{Eq:LLG}
  \der{}{\bhs}{t} + \alpha \, \bhs \times \der{}{\bhs}{t} = \TOR\, ,
\end{equation}
where $\alpha$ is the Gilbert damping parameter. The right-hand
side represents the torques due to  effective magnetic field and
spin-transfer,
\begin{equation}
\label{Eq:TOR}
  \TOR = -|\gyro| \mu_0 \, \bhs \times \Heff + \frac{|\gyro|}{\Ms d} \, \tor \, ,
\end{equation}
where $\gyro$ is the gyromagnetic ratio, $\mu_{0}$ is the vacuum
permeability, $\Ms$ stands for the saturation magnetization of the
free layer, and $d$ denotes thickness of the free layer.
Considering the thin ferromagnetic free layer of elliptical
cross-section, the effective magnetic field $\Heff$ can be written
as
\begin{equation}
\label{Eq:Heff}
  \Heff = -\hext \ez - \hani \left(\bhs \cdot \ez \right) \ez + \Hdemag\, ,
\end{equation}
and includes applied external magnetic field ($\hext$), uniaxial
anisotropy field ($\hani$), and the demagnetization field
($\Hdemag$), where $\ez$ is the unit vector along the $z$-axis
(easy axis), compare scheme in Fig.~\ref{Fig:DSV_scheme}. The
demagnetization field can be written in the form $\Hdemag = (\bhs
\cdot \DEMAG) \Ms = (\Hdx \sx, \Hdy \sy, \Hdz \sz)$, where
$\DEMAG$ is a diagonal demagnetization tensor.


\subsection*{Spin-transfer torque}

Generally, STT acting on a magnetic layer is determined by the
electron spin angular momentum absorbed from conduction electrons
within a few interfacial atomic layers of the ferromagnet
\cite{Stiles2002:PRB}. Thus, the STT acting on the central layer
$\FC$ can be calculated as $\tor = (\hbar / 2)\, ({\bf
j}_{\perp}^{\rm L} - {\bf j}_{\perp}^{\rm R})$, where ${\bf
j}_{\perp}^{\rm L}$ and ${\bf j}_{\perp}^{\rm R}$ are the
 spin current components perpendicular to magnetic moment of the free layer and
 calculated at the corresponding left and
right normal-metal/ferromagnet interfaces. The torque consists of
in-plane $\torip$ (in the plane formed by magnetic moments of the
two interacting films) and out-of-plane $\torop$ (normal to this
plane) parts, $\tor = \torip + \torop$, which have the following
forms
\begin{subequations}
\label{Eqs:STT}
  \begin{align}
  \label{Eq:STTpar}
    \torip &= I\, \bhs \times \left[ \bhs \times \left( \aL \SL - \aR \SR \right) \right]\, , \\
  \label{Eq:STTperp}
    \torop &= I\, \bhs \times \left( \bL \SL - \bR \SR \right)\, ,
  \end{align}
\end{subequations}
where $I$ is the current density, and $\SL$ and $\SR$ are the unit
vectors pointing along the fixed net-spins of the $\FL$ and $\FR$
layers, respectively. The parameters $\aL$, $\aR$, $\bL$, and
$\bR$ depend, generally, on the magnetic configuration and
material composition of the system, and have been calculated in
the diffusive transport limit \cite{Barnas2005:PRB}. According to
first-principles calculations of the mixing conductance
\cite{Xia2002:PRB}, the out-of-plane torque in metallic structures
is about two orders of magnitude smaller than the in-plane torque.
Although this component has a minor influence on critical
currents, it may influence dynamical regimes, so we include it for
completeness in our numerical calculations.

Let us consider now how the STT affects spin dynamics of the free
magnetic layer. We rewrite Eq.(\ref{Eq:LLG}) in spherical
coordinates $(\phi, \theta)$, which obey $\bhs = (\cos\phi
\sin\theta , \sin\phi \sin\theta , \cos\theta)$; see
Fig.~\ref{Fig:DSV_scheme}. Defining unit base vectors, $\ep = (\ez
\times \bhs) / \sin\theta$ and $\et = \ep \times \bhs$,
Eq.(\ref{Eq:LLG}) can be rewritten as
\begin{equation}
  \der{}{}{t}
  \begin{pmatrix}
    \phi\\
    \theta
  \end{pmatrix}
  =
  \frac{1}{1 + \alpha^2}
  \begin{pmatrix}
    \sin^{-1}\theta & -\alpha \sin^{-1}\theta\\
    \alpha & 1
  \end{pmatrix}
  \begin{pmatrix}
    \vp\\
    \vt
  \end{pmatrix}\, ,
\label{Eq:LLG2}
\end{equation}
where the overall torques $\vt$ and $\vp$, changing the angles
$\theta$ and $\phi$, respectively, read
\begin{widetext}
  \begin{subequations}
    \begin{align}
       \vt &= \TOR \cdot \et =
          -|\gyro| \mu_0 \, (\Hdx - \Hdy) \cos\phi \sin\phi \sin\theta +
          \frac{|\gyro|}{\Ms d} \, \tort\, , \\
       \vp &= \TOR \cdot \ep =
          -|\gyro| \mu_0 \, \left[
          \hext + (\hani + \Hdx \cos^2\phi + \Hdy \sin^2\phi - \Hdz) \cos\theta
                                    \right] \sin\theta +
          \frac{|\gyro|}{\Ms d} \, \torp\, .
    \end{align}
  \label{Eqs:vtvp}
  \end{subequations}
\end{widetext}
The first terms in Eqs.~(\ref{Eqs:vtvp}) describe the torques due
to  demagnetization and anisotropy fields of the free layer, while
$\tort = \tortip + \tortop$ and $\torp = \torpip + \torpop$ are
the corresponding components of the current-induced torque, which
originate from $\torip$ and $\torop$.

As we have already mentioned above, the main contribution to STT
comes from $\torip$. Assuming now that magnetic moment of the left
magnetic layer is fixed along the $z$-axis, $\SL = (0, 0, 1)$, and
magnetic moment of the right magnetic layer is rotated  by an
angle $\Omega$ from the $z$-axis and fixed in the layer's plane
(see Fig.~\ref{Fig:DSV_scheme}), $\SR = (0, \sin\Omega,
\cos\Omega)$, one finds
\begin{subequations}
\label{Eqs:tors_ip}
  \begin{align}
  \label{Eq:tort_ip}
    \tortip &= (\aL - \aR \cos\Omega) I \sin\theta
            + \aR I \sin\Omega \sin\phi \cos\theta\, , \\
  \label{Eq:torp_ip}
    \torpip &= \aR I \cos\phi \sin\Omega\, .
  \end{align}
\end{subequations}
The component $\tortip$ consists of two terms. The first one is
analogous to the term which describes STT in a SSV. However, its
amplitude  is now modulated due to the presence of $\FR$. In turn,
the second term in Eq.~(\ref{Eq:tort_ip}) is nonzero only in
noncollinear magnetic configurations. From Eq.~(\ref{Eq:torp_ip})
follows that $\torpip$ is also nonzero  in noncollinear
configurations, and only if the magnetization points out of the
layer's plane ($\phi \neq \pi / 2$). When magnetic moments of the
outer magnetic layers are parallel ($\Omega = 0$), $\tortip = (\aL
- \aR) I \sin\theta$. For a symmetric DSV, $\aL(\theta) =
\aR(\theta)$, and hence STT acting on $\bhs$ vanishes. On the
other hand, in the antiparallel configuration of $\SL$ and $\SR$
($\Omega = \pi$), the maximal spin torque enhancement can be
achieved,
 $\tortip = (\aL +\aR) I \sin\theta$.

Similar analysis of $\torop$ leads to the following formulas for
$\tortop$ and $\torpop$:
\begin{subequations}
\label{Eqs:tors_op}
  \begin{align}
  \label{Eq:tort_op}
    \tortop &= \bR I \cos\phi \sin\Omega\, , \\
  \label{Eq:torp_op}
    \torpop &= -(\bL - \bR \cos\Omega) I \sin\theta + \bR \cos\theta \sin\phi \sin\Omega\, .
  \end{align}
\end{subequations}
Thus, if the outer magnetic moments are collinear, $\tortop = 0$,
while $\torpop$ reduces to $\torpop = -(\bL - \bR) I \sin\theta$
for $\Omega = 0$, and $\torpop = -(\bL + \bR) I \sin\theta$ for
$\Omega = \pi$. Hence, in symmetric spin valves, where $\bL =
\bR$, $\torpop$ vanishes in the parallel configuration of the
outermost magnetic moments and is enhanced in the  antiparallel
configuration.

In the following sections we investigate STT and its effects on
critical current and spin dynamics. We start from symmetric spin
valves in antiparallel magnetic configuration ($\Omega = \pi$).
Then, we proceed with asymmetric exchange-biased dual spin valves.


\section{Symmetric DSVs}
\label{Sec:AntisymGeom}

Let us consider first symmetric DSVs with antiparallel orientation
of magnetic moments of the outermost ferromagnetic films: $\SL =
(0, 0, 1)$ and $\SR = (0, 0, -1)$. As indicated by
Eqs.~(\ref{Eqs:tors_ip}) and (\ref{Eqs:tors_op}), such a
configuration may lead to enhancement of STT in comparison to that
in SSVs. Thus, let us analyze first STT in two types of
structures: the double spin valve F(20)/Cu(10)/F(8)/Cu(10)/F(20)
in the antiparallel configuration, and the corresponding single
spin valve F(20)/Cu(10)/F(8). The numbers in brackets correspond
to layer thicknesses in nanometers.

\begin{figure}[h!]
  \centering
  \includegraphics[width=0.98\columnwidth]{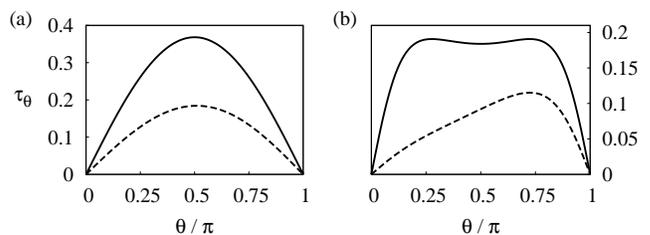}
  \caption{Spin transfer torque $\tort$ in symmetric DSVs,
  F(20)/Cu(10)/F(8)/Cu(10)/F(20), in the
  antiparallel configuration, $\Omega=\pi$, (solid lines),
           and  STT in SSVs, F(20)/Cu(10)/F(8) (dashed lines), where
           (a)~F = Permalloy, (b)~F = Cobalt.
           STT is shown in the units of $\hbar |I| / |e|$, and calculated
           for $\phi = \pi/2$. The material parameters as in Ref.\onlinecite{Gmitra2006:APL}.}
  \label{Fig:STT_sym}
\end{figure}

In Fig.~\ref{Fig:STT_sym} we show the angular dependence of STT
for DSVs and SSVs, when the vector $\bhs$ changes its orientation,
described by angle $\theta$, in the layer plane ($\phi = \pi/2$).
The magnetic layers made of Permalloy, Ni$_{80}$Fe$_{20}$
(Fig.~\ref{Fig:STT_sym}a), and of Cobalt (Fig.~\ref{Fig:STT_sym}b)
are considered. Due to the additional fixed layer ($\FR$), STT in
DSVs is about twice as large as in SSVs, which is consistent with
Berger's predictions~\cite{Berger1996:PRB}. Additionally, the
angular dependence of STT acting on the free layer in Co/Cu/Co
spin valves is more asymmetric than in Py/Cu/Py. This asymmetry,
however, disappears in Co/Cu/Co/Cu/Co DSVs due to superposition of
the contributions from both fixed magnetic layers to the STT.

The enhancement of STT in dual spin valves may lead to two
important improvements of the spin dynamics: reduction of the
critical current needed to trigger the spin dynamics, and decrease
of the switching time. The latter is defined as the time needed to
switch the magnetization from one stable position to the opposite
one. The fixed points of the dynamics of $\bhs$ are given by the
equations $\vt = 0$ and $\vp = 0$. If $\Omega = \pi$, they are
satisfied for $\theta = 0$, and $\theta =\pi$. Employing the
'zero-trace' stability condition of the linearized
Landau-Lifshitz-Gilbert equation\cite{Wiggins1990:Springer}, we
find the  critical current destabilizing the initial ($\theta =
0$) state in the form
\begin{equation}
  I^{0}_{c,\, {\rm DSV}} = \frac{\alpha \mu_0 \Ms d}{a^0_{\rm L} + a^0_{\rm R}}\,
                           \left(
                             \hext + \hani + \frac{\Hdx + \Hdy}{2} - \Hdz
                           \right)\, ,
  \label{Eq:IcAG_0}
\end{equation}
where $a^0_{\rm L}$ and $a^0_{\rm R}$ are calculated for $\theta
\rightarrow 0$. Additionally, we have omitted here the terms
resulting from $\torop$ because of their small contribution to the
critical current. Equation~(\ref{Eq:IcAG_0}) is analogous to the
expression for critical current in SSV~\cite{Gmitra2006:APL}.
According to our calculations, the critical current in DSVs with F
= Cobalt is $6$-times smaller than in SSVs, as reported by
Berger~\cite{Berger2003:JAP}. However, if F = Permalloy, the
introduction of a second fixed magnetic layer reduces the critical
current only by a factor of $2$. This difference arises from the
dependence of spin accumulation on spin-flip length which is about
$10$-times longer in Cobalt than in
Permalloy~\cite{Barnas2005:PRB}.

\begin{figure}[h!]
  \centering
  \includegraphics[width=0.99\columnwidth]{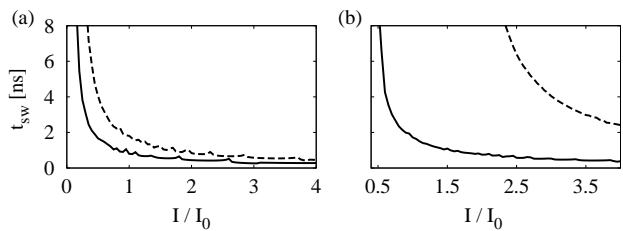}
  \caption{Switching time in DSVs F(20)/Cu(10)/F(8)/Cu(10)/F(20)
           in the antiparallel magnetic configuration, $\Omega=\pi$, (solid
           lines), and in SSVs F(20)/Cu(10)/F(8) (dashed lines), where
           (a)~F = Permalloy, (b)~F = Cobalt. The switching time
           is shown as a function of the normalized current density $I/I_0$, with
           $I_0 = 10^8 \Amp\cm^{-2}$. The other parameters as in Fig.2. }
  \label{Fig:tsw}
\end{figure}

To estimate the switching time in DSVs as well as in SSVs we
employ the single-domain macrospin approximation to the central
layer. Equations~(\ref{Eq:LLG}), including Eq.~(\ref{Eq:Heff}) and
Eq.~(\ref{Eqs:STT}), completely describe the dynamics of central
layer's spin, $\bhs$. In our simulations we assumed a  constant
current of density $I$. The positive current ($I > 0$) is defined
for electrons flowing from $\FR$ towards $\FL$ (current then flows
from $\FL$ towards $\FR$); opposite current is negative. Apart
from this, the demagnetization field of the free layer of
elliptical cross-section with the axes' lengths $130\,\nm$ and
$60\,\nm$ has been assumed, while external magnetic field was
excluded, $\hext = 0$. For each value of the current density, from
Eq.~(\ref{Eq:LLG}) we found evolution of $\bhs$ starting from an
initial state until $\bhs$ switched to the opposite state
(provided such a switching was admitted). In all simulation, the
spin was initially slightly tilted  in the layer plane from the
orientation $\bhs = (0, 0, 1)$, assuming $\theta_0 = 1^{\circ}$
and $\phi_0 = \pi/2$. A successful switching, with the switching
time $\ts$, is counted when $\ema (\ts) < -0.99$, where $\ema(t)$
is the exponentially weighted moving average~\cite{Roberts2000},
$\ema(t) = \eta~ \sz (t) + (1 - \eta) \ema (t - \Delta t)$,
$\Delta t$ is the integration step, and the weighting parameter
$\eta = 0.1$. The moving average $\ema$ is calculated only in time
intervals, where $\sz(t)$ remains continuously below the value of
$-0.9$; otherwise $\ema(t) = s_z(t)$. Fig.~\ref{Fig:tsw} compares
the switching times in DSVs and in the corresponding SSVs. In both
cases shown in Fig.~\ref{Fig:tsw}, a considerable reduction of the
switching time is observed in DSVs. Similarly as for the critical
current, the reduction of current required for switching in DSVs
with Cobalt layers is larger from that in DSVs with Permalloy
layers.


\section{Exchange-biased DSV}
\label{Sec:EBDSV}

Let us consider now an asymmetric exchange-biased DSV structure
with an antiferromagnetic layer IrMn adjacent to the $\FR$ layer
in order to pin its magnetic moment in a required orientation,
i.e. the structure Co(20)/Cu(10)/Py(4)/Cu(4)/Co(10)/IrMn(8). The
left magnetic layer, $\FL$ = Co(20), is assumed to be thick enough
so its magnetic moment is fixed, $\SL = \ez$. In turn, magnetic
moment of the right ferromagnetic layer, $\FR$ = Co(10), is fixed
in the layer plane at a certain angle $\Omega$ with respect to
$\SL$ due to the exchange-bias coupling to IrMn.

From Eqs.~(\ref{Eqs:vtvp}), (\ref{Eqs:tors_ip}), and
(\ref{Eqs:tors_op}) follows, that in a general case ($\Omega \neq
0$), the points $\theta = 0$ and $\theta =\pi$ are no more
solutions of the conditions for fixed points, $\vt = 0$ and $\vp =
0$, because of the additional terms in STT, which appear in
non-collinear situations (discussed in section II). These
additional terms lead to a nontrivial $\theta$-dependence  of STT,
and to a shift of the fixed points out of the collinear positions.

\begin{figure}[t!]
  \centering
  \includegraphics[width=0.98\columnwidth]{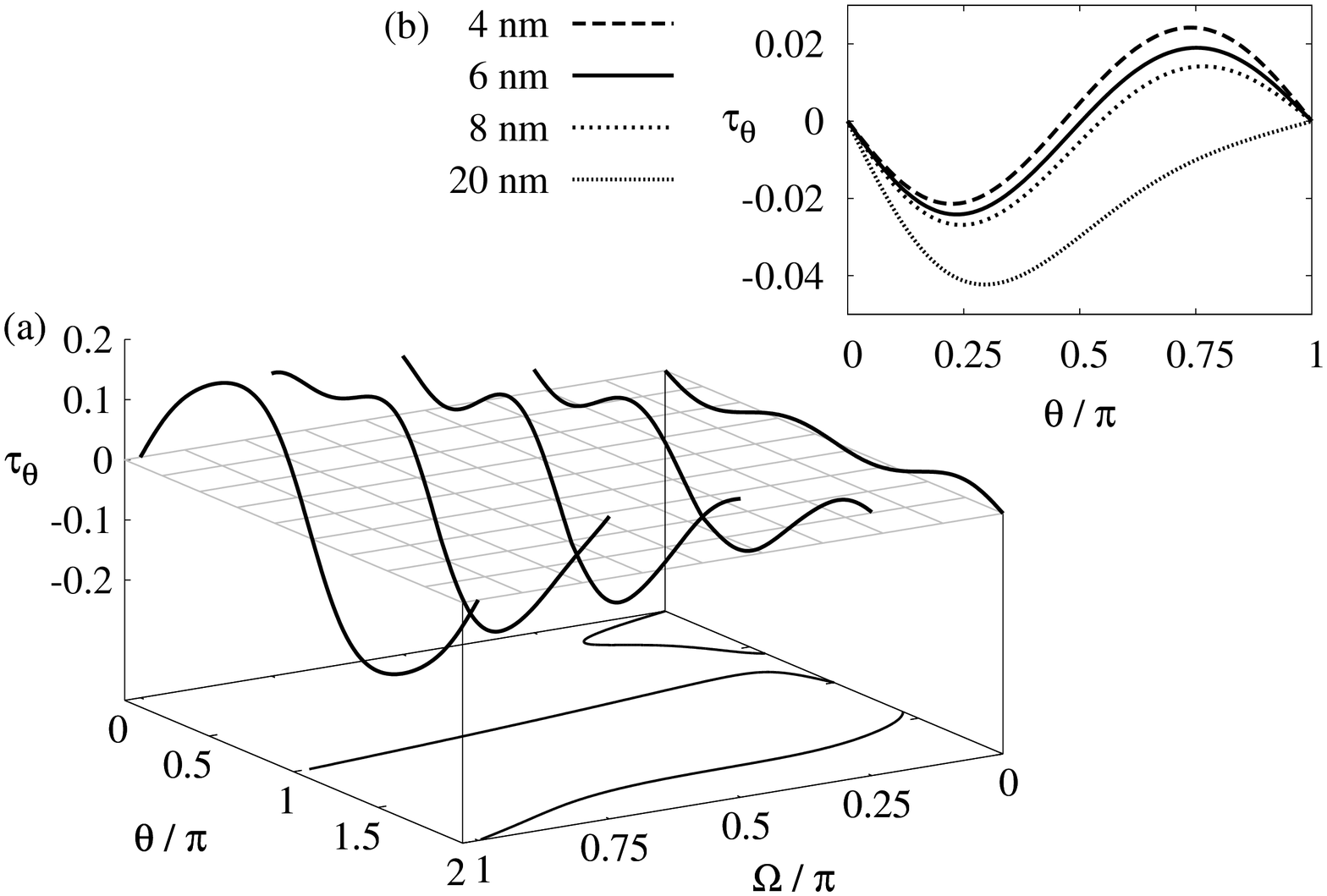}
  \caption{(a) STT acting on the central magnetic layer in
           Co(20)/Cu(10)/Py(4)/Cu(4)/Co(10)/IrMn(8)
           exchange-biased DSV
           as a function of the angle $\theta$, calculated for
           $\Omega = k \pi / 4,\, k = 0, 1, 2, 3, 4$.
           The contour plot in the base plane corresponds to zeros of $\tau_\theta$.
           (b) Wavy-like STT angular dependence in exchange biased DSV for $\Omega =
           0$, calculated for different thicknesses of the $\FR$ layer.
           STT is shown in the units of $\hbar |I| / |e|$. The other parameters as in Fig.2.}
  \label{Fig:STT_EBDSV}
\end{figure}

To analyze this effect in more details, let us consider first STT
assuming $\bhs$ in the layer plane ($\phi = \pi / 2$). According
to Eq.~(\ref{Eq:torp_ip}) and Eq.~(\ref{Eq:torp_op}), and to the
fact that the parameters $b$ are much smaller than $a$,  the
component $\torp$ is very small. In Fig.~\ref{Fig:STT_EBDSV}(a) we
show the second component of the torque, namely $\tau_\theta$, as
a function of the angle $\theta$ and for different values of the
angle $\Omega$. The configurations where $\tau_\theta = 0$ are
presented by the contour in the base plane of
Fig.~\ref{Fig:STT_EBDSV}(a).

In the whole range of the angle $\Omega$, two 'trivial' zero
points are present. Additionally, for small angles close to
$\Omega=0$, two additional zero points occur. The appearance of
these additional zero points closely resembles non-standard {\em
wavy}-like STT angular dependence, which has been already reported
in single spin valves with magnetic layers made of different
materials \cite{Barnas2005:PRB,Gmitra2006:PRL,Boulle2008:PRB}.
Such a $\theta$-dependence dictates that both zero-current fixed
points ($\theta=0,\pi$) become simultaneously stabilized
(destabilized) for positive (negative) current. This behavior is
of special importance for stabilization of the collinear
configurations, and for microwave generation driven  by current
only (without external magnetic field)
\cite{Gmitra2006:PRL,Barnas2005:PRB,Balaz2009:PRB}. We note, that
in contrast to SSVs, the {\em wavy}-like $\theta$-dependence in
exchange-biased DSVs is related rather to asymmetric geometry of
the multilayer than to bulk and interface spin asymmetries. This
trend is depicted in Fig.~\ref{Fig:STT_EBDSV}(b), where we show
variation of STT for different thicknesses of $\FR$ at $\Omega =
0$. The {\em wavy}-like torque angular dependence appears for the
thickness of $\FR$ markedly different from that of $\FL$.

Making use of the above described STT calculations, extended to
arbitrary orientation of $\bhs$, we performed numerical
simulations of spin dynamics induced by a constant current in zero
external magnetic field ($\hext = 0$). The sample cross-section
was assumed in the form of an ellipse with the axes' lengths
$130\,\nm$ and $60\,\nm$. As before, the numerical analysis has
been performed within the macrospin framework, integrating
equation of motion, Eq.(\ref{Eq:LLG}). In the simulation we
analyzed the long-term current-induced spin dynamics started from
the initial state corresponding to $\theta_0 = 1^{\circ}$, $\phi_0
= \pi / 2$.

\begin{figure}[t!]
  \centering
  \includegraphics[width=8cm]{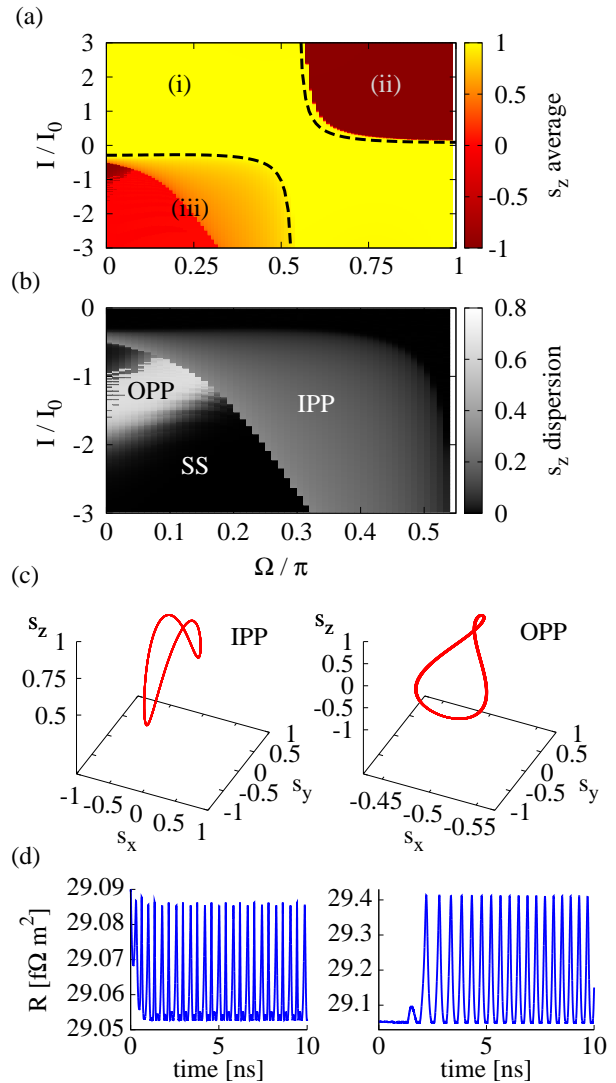}
  \caption{(Color online) Dynamical phase diagram for
           Co(20)/Cu(10)/Py(4)/Cu(4)/Co(10)/IrMn(8)
           exchange-biased double spin valve
           as a function of the angle $\Omega$ and normalized current
           density $I/I_0$ (with
           $I_0 = 10^8 \Amp\cm^{-2}$):
           (a) average value of the $\sz$ spin component;
           (b) dispersion of the $\sz$ spin component;
           (c) typical precessional orbits;
           (d) resistance oscillations associated with the IPP
               for $\Omega =0.4\pi$ and $I = -1.3 I_0$ (left part),
               and with the OPP for $\Omega = 0.1\pi$ and $I = -1.3 I_0$ (right part).
               The other parameters as in Fig.3.}
  \label{Fig:Dynam_maps}
\end{figure}

As one might expect, the current-induced spin dynamics of the free
layer depends on the angle $\Omega$, current density $I$, and on
current direction. To designate different regimes of STT-induced
spin dynamics, we constructed a dynamical phase diagram as a
function of current and the angle $\Omega$. The diagram shows the
average value $\av{s_{z}}$ of the $z$ component of the free layer
net-spin in a stable dynamical regime (see
Fig.~\ref{Fig:Dynam_maps}a), as well as its dispersion $D(s_{z}) =
\sqrt{\av{s_{z}^{2}} - \av{s_{z}}^{2}}$ (see
Fig.~\ref{Fig:Dynam_maps}b). The average value $\av{s_{z}}$
provides an information on the  spin orientation, whereas the
dispersion distinguishes between static states (for which $D=0$)
and steady precessional regimes (where $D>0$), in which the $z$
component is involved. For each point in the diagram, a separate
run from the initial biased state $\phi_0 = \pi/2$ and $\theta_0 =
1^{\circ}$ was performed. In the $\av{\sz}$ diagram,
Fig.~\ref{Fig:Dynam_maps}(a), one can distinguish three specific
regions. Region~(i) covers parameters for which a weak dynamics is
induced only: $\bhs$ finishes in the equilibrium stable point
which is very close to $\bhs = (0, 0, 1)$. As the angle $\Omega$
increases, STT becomes strong enough to cause switching, as
observed in the region~(ii). The higher the angle $\Omega$, the
smaller is the critical current needed for destabilization of the
initial state. For smaller $\Omega$, the current-induced dynamics
occurs for currents flowing in the opposite direction, see the
region~(iii). This behavior is caused by different sign of STT in
the initial state. From $\av{\sz}$ we conclude, that neither of
the previously mentioned stable states is reached. To elucidate
the dynamics in region~(iii), we constructed the corresponding map
of $D(\sz)$, Fig.~\ref{Fig:Dynam_maps}(b). This map reveals three
different modes of current-induced dynamics. For small current
amplitudes, in-plane precession (IPP) around the initial stable
position is observed. The precessional angle rises with increasing
current amplitude. Above a  certain critical value of $I$, the
precessions turn to out-of-plane precessions (OPPs). In a certain
range of $\Omega$, the OPPs collapse to a static state (SS), where
the spin $\bhs$ remains in an out-of-plane position close to
$\pm\ex$.

As $\SR$ departs from the collinear orientation, the critical
current needed for destabilization of the initial state increases.
This growth is mostly pronounced close to $\Omega = \pi/2$. To
describe the critical current, we analyzed Eq.~(\ref{Eq:LLG2})
with respect to the stability of $\bhs$ in the upper position.
Assuming that even in noncollinear configuration the stable
position of $\bhs$ is close to $\theta = 0$, we have linearized
Eq.~(\ref{Eq:LLG2}) around this point for arbitrary $\Omega$.
Then, for the critical current needed for destabilization of the
considered stable state we find
\begin{equation}
  I^{0}_{c,\, {\rm EBDSV}} \simeq \frac{\alpha \mu_0 \Ms d\,
                             \left(
                               \hani + \frac{\Hdx + \Hdy}{2} - \Hdz
                             \right)}
                             {a^{0}_{\rm L}(\Omega) - a^{0}_{\rm R}(\Omega) \cos\Omega}\, ,
  \label{Eq:Ic_0}
\end{equation}
where $a^{0}_{\rm L}(\Omega)$ and $a^{0}_{\rm R}(\Omega)$ are
calculated (for each configuration) assuming $\theta \rightarrow
0$. Comparison of Eq.~(\ref{Eq:Ic_0}) with the results of
numerical simulations is shown in Fig.~\ref{Fig:Dynam_maps}. When
considering the opposite stable point as the initial state, we
need to take $\aL$ and $\aR$ for $\theta \rightarrow \pi$.
Clearly, to destabilize the $\theta = \pi$ state one needs current
of opposite direction.

Current-induced oscillations are usually observed experimentally
{\it via} the magnetoresistance effect \cite{Kiselev2003:Nature}.
When electric current is constant, then magnetic oscillations
cause the corresponding resistance oscillations, which in turn
lead to voltage oscillations. The later are measured directly in
experiments. In Fig.~\ref{Fig:Dynam_maps}(d) we show oscillations in
the system resistance associated with the IPP (left) and OPP
(right). As the amplitude of the oscillations
corresponding to the OPP mode is sufficiently large to be measured
experimentally, the amplitude associated with the IPP mode is
relatively small. This is the reason, why IPP mode is usually not
seen in experiments.


\section{Discussion and conclusions}
\label{Sec:Conclusions}

We have calculated STT in metallic dual spin valves for arbitrary
magnetic configuration of the system, but with magnetic moments of
the outer magnetic films fixed in their planes either by large
coercieve fields or by exchange anisotropy. In the case of
symmetric DSV structures, we found a considerable enhancement of
STT in the antiparallel magnetic configuration. This torque
enhancement leads to reduction of the critical current for
switching as well as to reduction of the switching time. The
switching improvement has been found to be dependent on the
spin-flip lengths in the magnetic and nonmagnetic layers.
According to our numerical simulations, an ultrafast subnanosecond
current-induced switching processes can occur in DSVs with
antiparallel magnetic moments of the outermost magnetic films.

In exchange-biased spin valves we have identified conditions which
can lead to various types of spin dynamics. For $\Omega \lesssim
\pi / 2$, the negative current excites the central magnetic layer,
while for $\Omega \gtrsim \pi / 2$, opposite current direction is
needed. We have evaluated parameters for which switching to a new
stable state or to a precessional regime appears. This is
especially interesting from the application point of view.
However, the static state (SS) should be treated carefully. Such a
state is often connected with the {\em wavy}-like angular
dependence of
STT~\cite{Boulle2008:PRB,Gmitra2006:APL,Balaz2009:PRB}. The SS
becomes stable in the framework of macrospin model, when STT
disappears in a certain noncollinear configuration. However,
experimentally such a state has not been observed so
far~\cite{Boulle2008:PRB}. It has been shown in more realistic
micromagnetic approach, that this static state may correspond to
out-of-plane precessions~\cite{Jaromirska_2009}. The reason of
this is the finite strength of exchange coupling in magnetic film,
which does not fully comply with the macrospin approximation in
some cases.

Finally, we have derived an approximate formula for the critical
current in exchange biased DSVs, valid  for non-collinear magnetic
configurations. This formula,  Eq.(\ref{Eq:Ic_0}), is in good
agreement with the numerical simulations.

\subsection*{Acknowledgment}

The work has been supported by the the EU through the Marie Curie
Training network SPINSWITCH (MRTN-CT-2006-035327). M.G. also
acknowledges support VEGA under Grant No. 1/0128/08, and J.B.
acknowledges support from the Ministry of Science and Higher
Education as a research project in years 2006-2009.



\end{document}